\pdfoutput=1
\documentclass[twocolumn,showpacs,preprintnumbers,amsmath,amssymb]{revtex4}

\usepackage{xspace}
\usepackage{epsfig}
\usepackage{amsfonts}
\usepackage{graphicx}
\usepackage{url}
\usepackage[colorlinks,%
          linkcolor=blue,%
        citecolor = blue,%
              hyperindex,%
        plainpages=false,%
         urlcolor = blue,%
       pdfstartview=FitH]{hyperref}

\begin{document}

\title{Polarization-Engineering in III-V Nitride Heterostructures:\\ New Opportunities For Device Design}


\author{%
  Debdeep Jena (djena@nd.edu), 
  John Simon,
  Albert (Kejia) Wang,
  Yu Cao,
  Kevin Goodman,
  Jai Verma,
  Satyaki Ganguly,
  Guowang Li,
  Kamal Karda,
  Vladimir Protasenko,
  Chuanxin Lian,
  Thomas Kosel,
  Patrick Fay,
  Huili Xing}
  

\affiliation{Electrical Engineering, University of Notre Dame, IN, 46556, USA}

\date{\today}

\begin{abstract}
The role of spontaneous and piezoelectric polarization in III-V nitride heterostructure devices is discussed.  Problems as well as opportunities in incorporating polarization in abrupt and graded heterojunctions composed of binary, ternary, and quaternary nitrides are outlined.
\end{abstract}

\maketitle

\section{Introduction}
Spontaneous polarization exists along the $(0001)$ (metal-polar) and $(000\bar{1})$ (N-polar) directions of wurtzite III-V nitride semiconductor crystals.  Strain-induced piezoelectric polarization can either add to, or subtract from the spontaneous polarization in heterostructures based on the composition and polarity.  At a sharp heterojunction, the discontinuity in polarization leads to the appearance of a {\em immobile} polarization sheet charge (unit: C/cm$^2$) of density
\begin{equation}
\label{eq1}
\sigma_{\pi} = ({\bf P_{1} - P_{2}}) \cdot \hat{\bf n},
\end{equation}
where ${\bf P_{1}, P_{2}}$ are the total (spontaneous + piezoelectric) polarization vectors across the junction, and $\hat{\bf n}$ is the unit normal to the heterointerface plane. If the heterojunction is compositionally graded (say along the growth direction $z$) instead, Eq.~\ref{eq1} is modified to its 3D counterpart
\begin{equation}
\label{graded}
\rho_{\pi}(z) =  - \nabla \cdot {\bf P(r)} = - \frac{\partial P_{z}(z)}{\partial z},
\end{equation}
where $\rho_{\pi}(z)$ is the {\em immobile} 3D bulk charge density (unit: C/cm$^{3}$).  The built-in {\em real} electric fields originating from the polarization charges, when combined with the {\em quasi}-electric fields at sharp and graded heterojunctions lead to a rich range of phenomena that impact every conceivable wurtzite nitride heterostructure device.  This work reviews the existing body of knowledge on the device consequences of the high polarization fields, and outlines several future possibilities that exploit as-yet  unrealized applications of polarization engineering.  The materials challenges in realizing the proposed novel applications are mentioned.  Though structures grown along semi-polar directions are not discussed explicitly, similar arguments are expected to apply.


\section{Polarization in Al$_x$In$_y$Ga$_{1-x-y}$N alloys}

\begin{figure*}[htb]%
\includegraphics*[width=\textwidth]{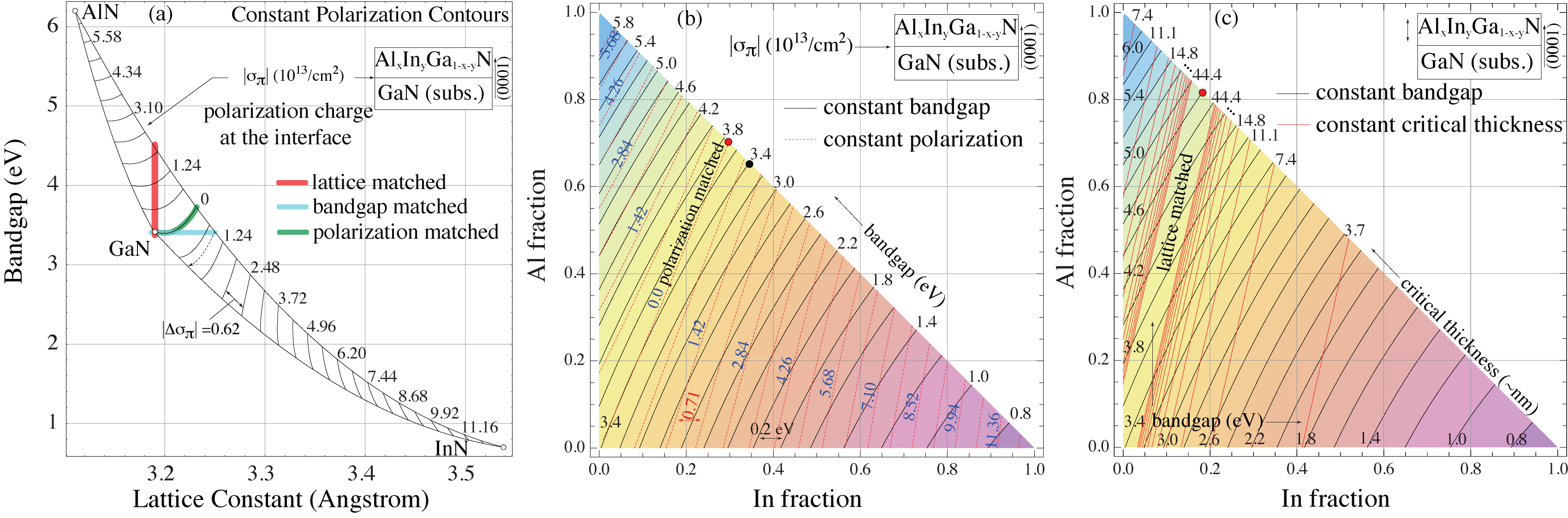}
\caption{ (a) Constant polarization (spontaneous + piezoelectric) contours on the energy bandgap-lattice constant plot for nitrides grown lattice matched to GaN along the (0001) orientation.  The lattice-matched, bandgap-matched, and polarization-matched compositions are highlighted. (b) Contours of constant polarization (red dashed lines) and constant bandgap (black solid lines) as a function of the In and Al mole fractions in AlInGaN layers grown strained on (0001) oriented GaN. (c) Contours of constant critical thickness (red solid lines) and constant bandgap (black solid lines) as a function of the In and Al mole fractions in AlInGaN layers grown strained on (0001) oriented GaN.
}
\label{fig1}
\end{figure*}

Fig.~\ref{fig1}(a) shows the contours of constant polarization sheet charge at Al$_x$In$_y$Ga$_{1-x-y}$N/GaN heterojunctions grown coherently strained on (0001) oriented relaxed GaN substrates.  Bowing parameters of $b=-1.0, -2.0, -4.0$ eV for AlGaN, InGaN, and InAlN ternaries are used, and quaternary material parameters are taken from \cite{bernardiniBook}.  We note that these are representative values, and in particular, parameters for InAlN are currently under investigation.  The polarization coefficients used are taken from \cite{bernardiniBook}.  Recent investigations of the growth and applications of InAlN and InAlGaN layers lattice matched to GaN (falling on the red line) have yielded very encouraging results for both electronic and optical devices, greatly expanding the range of material choices available for III-V nitride devices.  The rapidly developing growth technology - by both MOCVD and MBE makes the realization of quaternaries and InAlN ternaries with a range of compositions a tantalizing prospect.  When such technology is at hand, heterostructures with compositions that are polarization-matched (green line), and/or bandgap matched (blue line) can help open a number of new applications.  These materials offer a potential method of solving existing problems encountered in typical Al(Ga)N/GaN and In(Ga)N/GaN - containing heterostructures that form most devices today.  To aid crystal growers, Fig.~\ref{fig1}(a) is recast in Fig.~\ref{fig1}(b) which shows the bandgap and polarization dependence on the Al and In mole fractions in Al$_x$In$_y$Ga$_{1-x-y}$N grown on GaN.  

Strain in Al$_x$In$_y$Ga$_{1-x-y}$N alloys limits the thickness that can be grown coherently on GaN.  A crude estimate of the critical thicknesses based on Blanc's estimate may be obtained: $t_{cr} \sim b_{e}/2 \epsilon(x,y)$, where $b_{e}$ is the Burgers vector, and $\epsilon(x,y)$ is the strain.  Fig.~\ref{fig1}(c) shows the estimated thicknesses that can be grown on GaN coherently; we note here that the actual coherent layer thicknesses can vary, and the numbers in the figure should only be taken as a rough estimate at best.

\section{Polarization effects in existing devices}

The high 2DEG charges in conventional Al(GaIn)N/GaN high electron mobility transistors (HEMTs) are entirely polarization induced, which enable high drive currents \cite{morkoc_polar_book}.  Coupled with the high breakdown voltages originating from the large bandgap, nitride HEMTs have proven very attractive for high-frequency, high-power electronic device applications.  Strain in the Al(Ga)N barrier layers has been identified as a possible source of device degradation under stress, and recent reports indicate that lattice-matched InAlN barriers may prove more robust.  Lattice-matched ternary \cite{grandjean06apl} and quaternary \cite{lim10apl} barrier HEMTs have also been reported recently, showing promising performance benefits over low Al composition AlGaN barrier layers.  Polarization-matched heterostructures are yet to be investigated for E-Mode HEMTs, but will require the source/drain and access regions to be populated by carriers similar to what was done recently for such devices grown along the non-polar direction \cite{fujiwara10}.

In optical devices, the role of polarization has been investigated in the active regions where the built-in electric field lowers the oscillator strength by reducing the electron-hole wavefunction overlap \cite{grandjean_polar_book}, the quantum-confined Stark effect (QCSE).  Under high carrier injection conditions, the polarization field is partially screened, but it causes spectral shift of emission with bias, and may prove problematic for ultra low threshold light emitters.  Though light emitting diodes and lasers over a large spectral window have been demonstrated with heterostructures grown along the polar directions, non- and semi-polar orientations are being intensively investigated recently \cite{speckMRS09}.

\section{Sharp Heterojunctions}

\subsection{Tunable quasi band-offsets}

\begin{figure}[t]%
\includegraphics*[width=\linewidth]{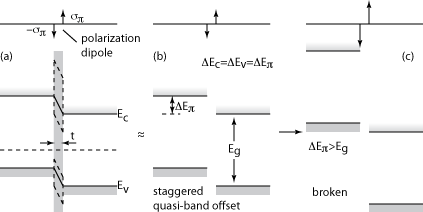}
\caption{%
  An ultrathin polarization dipole layer results in a quasi band-offset leading to staggered or broken gap alignments.}
\label{fig2}
\end{figure}

Consider an ultrathin layer (thickness= $t$) of polarization-mismatched nitride layer inserted between thick GaN layers shown in Fig.~\ref{fig2}(a).  The dipole layer could be of wider, narrower, or the same bandgap as the cladding layers with proper choice of composition and polarity; the cases are indicated schematically in Fig.~\ref{fig2}(a).  The polarization discontinuities of Eq.\ref{eq1} result in a charge dipole at the heterojunction with sheet charges $\pm \sigma_{\pi}$.  The electric field in the thin layer due to the polarization charges is $F_{\pi} = \sigma_{\pi}/\epsilon_{s}$, and the corresponding energy drop is $\Delta E_{\pi} = q F_{\pi} t = q \sigma_{\pi}t/\epsilon_{s}$, where $\epsilon_{s}$ is the dielectric constant of the semiconductor.  For example, if $t \sim 1$nm, and $F_{\pi} \sim 1$ MV/cm, then $\Delta E_{\pi} \sim 0.1$ eV, which is substantial.  This is achievable with, for example (but not limited to), a thin layer of AlInN that is {\em bandgap-matched} to GaN, which we assume does not introduce any chemical band offsets.  The value of $\Delta E_{\pi}$ can be much larger given the large polarization fields in nitrides.  

Note that an ultrathin polarization dipole layer can be effectively transparent to electron transport in the vertical direction, and therefore the dipole layer effectively mimics a {\em band offset}.  The magnitude of the band offset is equal in both the conduction and valence bands ($\Delta E_{c\pi} = \Delta E_{v\pi} = \Delta E_{\pi}$) and is therefore of the {\em staggered} type if the polarization-induced band offset $\Delta E_{\pi} < E_{g}$ is less than the bandgap ($E_{g}$) of the thicker cladding layers [Fig.~\ref{fig2}(b)].  Since such a band offset changes as $t$ and $\sigma_{\pi}$ are varied, it is {\em tunable}; indeed, it may even be possible to achieve an effectively {\em broken} gap offset if $\Delta E_{\pi} > E_{g}$ as shown in Fig.~\ref{fig2}(c).  The origin of this polarization-dipole induced band offset is a {\em real} electric field and therefore it acts equally on electrons and holes, giving the symmetry in the CB and VB offsets.  This is in stark contrast to a {\em quasi} electric field at heterojunctions, which is born out of changes in the chemical nature of constituent atoms and thereby can act very differently on electrons and holes leading to $\Delta E_{c} \neq \Delta E_{v}$ in general, and resulting in all three possible (straddling, staggered, and broken) band offsets \cite{kroemer}.  

Such artificial band-offsets have been investigated in GaN/AlN/GaN junctions \cite{stacia02}, \cite{caoyu10} and in GaN/InGaN/GaN junctions \cite{tomas06}.  The ultrathin layer for obtaining a polarization induced band offset may be of a smaller bandgap than the cladding layers for high transparency to charge transport in the growth direction, in which case it may require the expulsion of bound states in the thin asymmetric quantum wells to prevent charge trapping.  The ultrathin dipole layer could also be of a larger bandgap which will reduce the current density by requiring tunneling/thermionic emission.  By the proper choice of compositions, such polarization induced band offsets can be introduced {\em in conjunction with real band offsets} to increase, reduce, or cancel $\Delta E_{c}$ (or $\Delta E_{v}$) selectively.  The added degree of freedom provided by the large polarization in the III-V nitrides can prove to be a powerful device design tool.  

\subsection{Coupled electron-hole gases}

Consider an AlN layer of thickness $t$ embedded deep inside an undoped GaN bulk layer shown in Fig.~\ref{fig3}(a).  A critical thickness $t_{cr} = E_{g}/F_{\pi}$ is easily identified - if the dipole layer is thicker, the potential drop exceeds the bandgap of the cladding layers ($\Delta E_{\pi} > E_{g}$), resulting in an effectively broken-gap offset, whereby the valence band electron states are aligned with the empty conduction band states on the other side.  In a nominally undoped structure, assisted by the large polarization field, there is a finite probability for valence band electrons on the left to {\em tunnel} into the empty conduction band states on the right, leaving behind holes as indicated in Fig.~\ref{fig3}(a).  A rough estimate of the `time constant' involved in this process can be obtained.  The effective velocity of carriers in the valence band is $v \sim \hbar g / m_{0}$, where $g \sim \pi / a_{0}$ is a reciprocal lattice vector, $a_{0}$ is a lattice constant, and $m_{0}$ is the free electron mass.  An electron confined to a region of length $l_{0}$ collides the forbidden gap barrier at a rate $\gamma \sim v/l_{0}$, and the probability of tunneling is $T \approx \exp{(-2S)}$, where $S = \int \kappa(z) dz$ is the action, with $\kappa^{2}(z) = 2m [V(z) - E] / \hbar^{2}$.  The time for `escape' is estimated as $\tau \sim 1/\gamma T$, which for relevant GaN parameters evaluates to the order of seconds.

After such a process of tunneling relaxation has occured, the potential difference far from the junction region is $\Delta \phi_{\pi} = (\sigma_{\pi} - qn_{s}) t / \epsilon_{s}$ as shown in the energy band diagram in Fig.~\ref{fig3}(b).  Here $n_{s}$ is the sheet charge density (units: cm$^{-2}$) of a {\em mobile} electron gas, equal to the mobile hole charge density at the other interface.  The mobile charge dipole forms in response to the potential drop, in an attempt to screen the polarization dipole $\sigma_{\pi}$.  We note that charge neutrality requires the net charge to be zero, but the mobile charge dipole need not be equal to the polarization dipole.  This novel mechanism of formation of mobile carriers may also be pictured as `modulation-doping'.  However, for example for the 2DEG, the source of electrons are not donor impurities in a wider bandgap region, but the {\em valence band} on the other side of the heterojunction barrier.  This is a novel mechanism for creating mobile electrons and holes {\em without impurity doping}, and understanding it helps us exploit polarization in powerful ways.  

When the 2DEG and 2DHG form as indicated in Fig.~\ref{fig3}(b), the Coulombic coupling between them can be tuned by changing $t$.  A major challenge in {\em observing} such coupled electron-hole gas is forming separate electrical contacts to them.  Such closely coupled 2DEG and 2DHG assemblies are desirable for fundamental studies of Coulomb drag, and with suitable choice of the materials (and their bandstructures) can enable tunneling field-effect transistors of both of the `classical' type \cite{qin06_tfet_proposal}, and perhaps also of the phase-coherent many-particle type \cite{bisfet}.

\subsection{Resonant- and Zener-Tunneling Structures}

\begin{figure}[t]%
\includegraphics*[width=\linewidth]{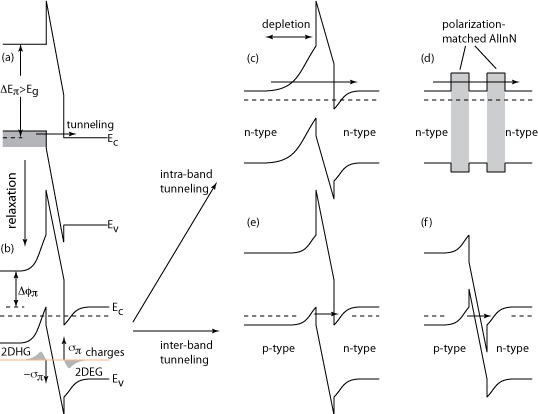}
\caption{%
  Various intraband and interband tunneling structures and the role of polarization.}
\label{fig3}
\end{figure}

Tunneling structures are best understood in polar III-V nitride heterostructures by building upon the nominally undoped structure shown in Fig.~\ref{fig3}(a, b) and discussed in the earlier section.  When the structure is uniformly doped (say n-type), the energy band diagram is as shown in Fig.~\ref{fig3}(c).  The heterojunction side that has a positive polarization sheet charge ($+\sigma_{\pi}$) sees an accumulation layer of electrons, whereas the other side, the role played by the mobile holes in the undoped structure is taken up by the immobile ionized donors.  If two such barriers are arranged to enable resonant tunneling, polarization essentially increases the tunneling distance by the depletion region thickness and leads to a lowering of the net tunneling current drive over a polarization-free structure.  This problem can be remedied by growing along non-polar orientations, or even when the growth is along the polar orientation, by choosing {\em polarization-matched} barriers (for example by using Al$_{0.7}$In$_{0.3}$N barriers as indicated in Fig.~\ref{fig3}(d) and Figs.~\ref{fig1}(a,b), though the barrier height may be restricted).  

If instead of uniform doping, the polarization dipole layer is surrounded by a p-n junction, the extra potential drop due to the dipole can either assist, or oppose the natural built-in potential of the p-n junction.  In the case where it opposes the built-in potential, it causes an extension of the depletion regions and is akin to a built-in reverse bias, and can be used for tuning the turn-on voltage and/or the reverse-bias breakdown voltage.  The opposite case is when the polarization dipole {\em assists} the dipole formed by ionized dopants by virtue of being in the same direction.  Is such a case, the polarization dipole shrinks the depletion region thickness and is similar to a built-in forward bias.  If the polarization charges are greater than necessary for dropping the entire built-in potential of the control p-n junction ($|\sigma_{\pi}|>q N_{d} x_{n} = q N_{a} x_{p}$), then {\em accumulation} regions form at the two heterojunctions, and the effective band offset between the two sides is now similar to a {\em broken-gap} heterojunction.  Such situations are shown in the energy band diagrams of Figs.~\ref{fig3}(e and f).  This arrangement enables Zener tunneling of electrons from the valence band of the p-side to the empty conduction band states on the n-side under reverse bias voltage, whereas no current flows under small forward bias since the valence band states are aligned with the bandgap of the n-side.  Thus the structure acts as a `backward diode'.  Such polarization-enabled Zener tunnel junctions have been recently experimentally demonstrated using GaN/AlN/GaN \cite{simon09}, \cite{grundmann07} and GaN/InGaN/GaN heterostructures \cite{sid10}, and carefully modeled \cite{schubert10prb}.  They hold much promise for future applications in wide-bandgap semiconductor devices.

Such interband tunneling diodes in traditional narrow-gap semiconductors are achievable primarily by heavy impurity doping with associated problems of discrete dopant effects and bandgap shrinkage. In nitride heterostructures the role of dopants is taken up by polarization charges that are confined to atomic planes, and can be far higher than what can be achieved by impurity doping.  Furthermore, the dipole layer may be chosen to be a narrower bandgap than the surrounding layers as in Fig.~\ref{fig3}(f), and thus the choice of polarization and composition offers a clear path for boosting the current drive.  Such polarization-induced Zener tunnel junctions can prove to be useful in zero-bias rectifiers, for electrically connecting multijunction solar cells and multicolor light emitters \cite{grundmann07}, and also possibly in tunneling field-effect transistors (TFETs) in the future.  

The large built-in potential drop can enable tunneling-enhanced Ohmic contacts.  Low-resistance Ohmic contacts to p-type wide-bandgaps has proven difficult due to the unavailability of very high work-function ($q\phi_{M}$) metals.  The resulting misalignment of the Fermi level of the metal with the valence band results in high barriers ($q \phi_{B} = q \chi_{s} + E_{g} - q \phi_{M}$).  This barrier can be lowered by an amount $\Delta E_{\pi} = q F_{\pi} t$ by introducing a polarization dipole layer.  The high electric field, and a narrower bandgap dipole layer can enable low-resistance contacts that are also reasonably optically transparent, as proposed in \cite{schubert10apl}.

\subsection{Polarization-balanced Electronic Devices}

The {\em precise equality} of the positive and negative components in a polarization dipole at symmetric nitride heterojunctions has led to the concept of `Natural Super-Junctions' (NSJs) for high-voltage applications.  Such devices when made using p-n junctions (for example in Silicon) by impurity doping place stringent requirements on the doping densities of donors and acceptors so that the vertical field is entirely supported by ionized dopants.  This requirement is naturally met in III-V nitride heterojunctions, and in addition the high density (and presumably high mobility) polarization-induced 2DEGs and 2DHGs can contribute to low on resistances \cite{panasonic08}.  Back-barriers for strong electron confinement is required for highly scaled GaN HEMTs.  Typical double heterostructure designs employ AlGaN back barriers for metal-polar devices, which causes a degree of reverse polarization and depletion of the 2DEG densities, causing higher sheet resistances.  One possibility for reducing this loss of charge is by using an AlInN layer that is polarization-matched to GaN.

\subsection{Polarization-balanced Optical Devices}

\begin{figure}[t]%
\begin{center}
\includegraphics*[width=\linewidth]{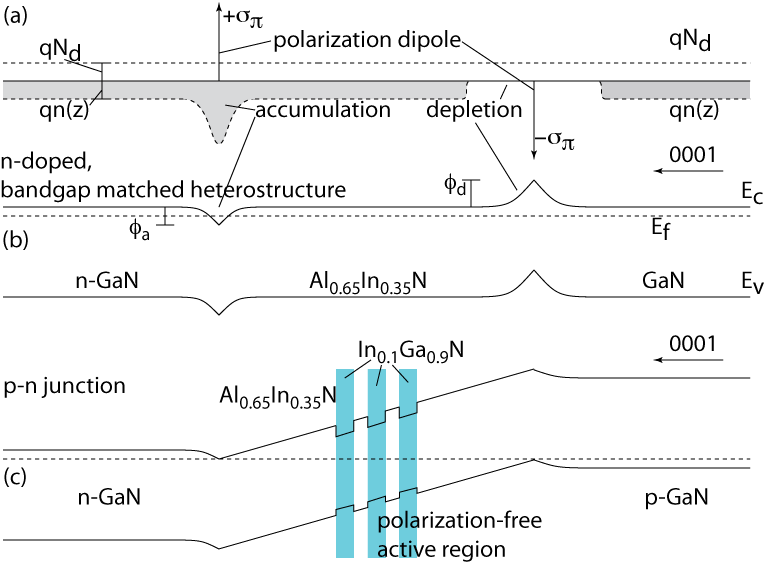}
\end{center}
\caption{%
  Schematic charge and energy band diagrams indicating electrostatics and polarization-matching.}
\label{fig4}
\end{figure}

Though growth along non-polar or semi-polar directions helps reduce QCSE based reduction of oscillator strengths by removing the built-in polarization fields, it is possible to do so by growing along the polar directions by choosing polarization-matched compositions \cite{schubert08polmatch}.  We discuss one possibility as an example, along with a few device electrostatics arguments.  Consider the case of a bandgap-matched GaN-Al$_{0.65}$In$_{0.35}$N heterostructure grown along the (0001) direction.  Polarization sheet charges $\pm \sigma_{\pi}$ form at the two heterojunctions without a band offset as indicated in Fig.~\ref{fig4}(a, b).  If the entire structure is uniformly doped with donor density $N_{d}$, the polarization charges cause depletion and accumulation of mobile electrons as indicated.  The peak electric fields at the heterojunction is $F_{max} = \sigma_{\pi}/2 \epsilon_{s} = F_{\pi}/2$ where $\epsilon_{s}$ is the dielectric constant.  The potential barrier height at the depleted junction is $\phi_{d} = \sigma_{\pi}^{2}/8 \epsilon_{s}qN_{d}$, and the potential `well' depth in the accumulation side is $\phi_{a} \approx F_{max} L_{D}$, where $L_{D} = \sqrt{\epsilon_{s} k T/ q^{2}N_{d}}$ is the Debye length.  Accumulation regions do not impede the motion of free carriers; thus, if a p-n junction structure as shown in Fig.~\ref{fig4}(c) is grown, accumulation regions form at both the heterojunctions.  If one chooses InGaN composition on the {\em same polarization contour} as the AlInN region for quantum wells, the active region is freed of polarization fields.  The material choices are easily seen in Fig.~\ref{fig1}(a or b): move from GaN$\rightarrow$InAlN along the bandgap-matched line, then move to the InGaN composition staying along the constant polarization contour (dashed line).   The materials challenges in doing so are indeed high, and it may be more prudent to free active regions from polarization by simply growing along non-polar orientations.  But the fundamental physics does not rule out the possibility of achieving polarization-free active regions even if the underlying structure is inherently polar.  In cases where the injection efficiency of holes (and/or electrons) are low due to deep dopant levels, similar polarization-matching approaches could prove highly attractive when combined with the added advantage of higher injection efficiency in structures grown along the polar orientation.  The possibility of polarization-enhanced p-type doping can be integrated into the device design along polar directions in graded bandgap heterostructures, which is discussed next.

\section{Graded Heterojunctions}

\begin{figure}[t]%
\includegraphics*[width=\linewidth]{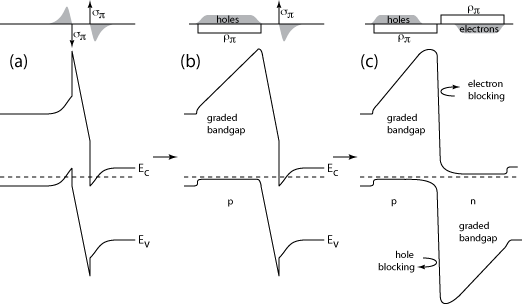}
\caption{%
  Energy band diagrams illustrating the role of polarization in graded bandgap structures.}
\label{fig5}
\end{figure}

When one or both polar heterojunctions are {\em compositionally graded}, a number of additional utilities of polarization emerge.  Consider the canonical structure of the last section as shown in Fig.~\ref{fig5}(a), and assume we keep one junction abrupt and compositionally grade the other as shown in Fig.~\ref{fig5}(b).  In that case, the {\em fixed bulk} polarization charge density in the graded region is given by Eq.~\ref{graded}:  $\rho_{\pi} = - \nabla \cdot {\bf P(r)}$.  In response to the macroscopic electric field created by the fixed polarization charge, neutralizing {\em mobile} charges on the graded side also spread from a 2D form to a 3D bulk form to screen/cancel the electric field.  We note here that the neutralizing charge may be localized in the presence of a high density of defects and traps, but can {\em always} be made mobile by compensation doping of the traps and defect states.  

\subsection{Optoelectronic devices: doping \& contacts}

Highly conductive p-type layers have been difficult to achieve in GaN based heterostructures owing to the high activation energy of Mg acceptors ($E_{A} \sim 180$ meV), leading to associated problems with hole injection efficiency in optical devices.  In addition, both Mg acceptor levels as well as Si donor levels ($E_{D}$) become deeper with increasing bandgap of Al(Ga)N layers that are useful for UV optoelectronics \cite{taniyasuNat06}, \cite{asifNatPhotonics08}.  When the activation energies are much larger than the thermal energy $(E_{A}, E_{D} >> kT)$, the low thermal ionization efficiency results in high resistivity layers.  The polarization-induced charge in short-period superlattices offers a method to {\em field-ionize} these deep dopants, and extract mobile carriers.  However, in such structures there still exist band offsets that impede carrier motion in the direction of most interest.  Interestingly, just as a buried polarization dipole results in a mobile electron-hole gas without impurity doping, a defect/trap-free compositionally graded layer should necessarily result in mobile carriers to screen the macroscopic electric field.  The role of dopants ($N_{A}$) in compositionally graded layers is reduced to {\em compensating} the trap states ($N_{T}$) that may localize the mobile charges, and maintaining the Fermi level close to the valence band (for p-type) or the conduction band (for n-type).  Thus the {\em mobile} carrier concentration $p$ is determined not by the impurity doping, but by the polarization charge density by the charge neutrality condition: $\rho_{\pi} + q N_{A}^{-} \approx q(p + N_{T}^{+})$, whereby the hole density is $p \approx \rho_{\pi}/q + (N_{A}^{-} - N_{T}^{+})$.  If complete compensation of the traps are achieved, $p \approx \rho_{\pi}/q$.  This form of polarization-induced doping was demonstrated for conductive n-type layers \cite{dj02apl}, and very recently for p-type layers, and has been incorporated into prototype UV LED structures \cite{simonSci10}.  An embodiment of a polarization-doped p-n junction (with $n \approx \rho_{\pi}/q + (N_{D}^{+} - N_{T}^{-})$) is shown in Fig.~\ref{fig5}(c).

A number of added features of such polarization-induced doping may be identified.  The graded bandgap layer can be of a larger bandgap and thus optically transparent for the active layer wavelengths.  It can enable natural optical mode confinement in the correct orientation, similar to graded-index separate confinement heterostructure (GRINSCH) lasers.  If the graded layer is p-type (a similar argument holds for n-type), $E_{F}- E_{v}$ is nearly constant, and the entire difference in the bandgap appears in the conduction band $E_{c}(z_{1}) - E_{c}(z_{2})= \Delta E_{g}$, leading to a quasi-electric field for {\em electrons}.  With the proper magnitude and orientation, this barrier can act as an electron blocking layer as shown in Fig.~\ref{fig5}(c).  In abrupt heterostructures with straddling band offsets, a conduction band offset $\Delta E_{c}$ based electron blocking layer necessarily comes with a valence band offset $\Delta E_{v} = \Delta E_{g} - \Delta E_{c}$, which results in hole blocking to a certain degree.  The only way to obtain a barrier for electrons and not for holes is to use a p-type doped graded-bandgap layer, where the {\em entire} bandgap difference appears in the conduction band, whereas the valence band is left without any barriers impeding hole flow.  This feature is well known in heterostructure bipolar transistors, where such doping can prove specially attractive.

\subsection{Heterostructure bipolar transistors}

The high resistance of p-type layers has proven to be the bottleneck in achieving high-performance heterostructure bipolar transistors (HBTs) using GaN.  Significant improvements in GaN HBTs have been made recently by incorporating narrower bandgap InGaN base layers.  The polarization-induced p-type doped graded gap layers described in the last section can help accelerate the progress in GaN HBTs \cite{asbeck00}.  Such p-type layers allow for wide-bandgap base and collector layers to take full advantage of the high breakdown of wide bandgaps, and simultaneously offer large built-in quasi-electric fields in the base to reduce the base transit time and improve the speed of such devices.

\section{Summary}

In summary, the recent expansion of GaN material research into quaternaries opens up new opportunities for polarization-based heterostructure device design.  A sampling was discussed here, and is expected to be extended in the future.

Acknowledgement: The authors acknowledge financial assistance from DARPA, ONR, NSF, and AFOSR.

%
%

\end{document}